
\magnification=\magstep1
\parskip 5mm plus 1mm

\vsize=7.5in
\hsize=5.6in
\tolerance 10000

\baselineskip 12pt plus 1pt minus 1pt
\pageno=0
\centerline{{\bf Validity of the Eikonal Approximation}\footnote{*}{This work
is
supported in part by funds
provided by the U. S. Department of Energy (D.O.E.) under contract
\#DE-AC02-76ER03069.}}
\vskip 24pt
\centerline{Daniel Kabat}
\vskip 12pt
\centerline{\it Center for Theoretical Physics}
\centerline{\it Laboratory for Nuclear Science}
\centerline{\it and Department of Physics}
\centerline{\it Massachusetts Institute of Technology}
\centerline{\it Cambridge, Massachusetts\ \ 02139\ \ \ U.S.A.}
\vskip 1.5in
\centerline{Submitted to: {\it Comments on Nuclear and Particle Physics}}
\vfill
\noindent CTP\#2095  \hfill April 1992
\eject
\baselineskip 24pt plus 2pt minus 2pt
\centerline{\bf ABSTRACT}
\medskip
We summarize results on the reliability of the eikonal approximation in
obtaining the high energy behavior of a two particle forward scattering
amplitude.  Reliability depends on the spin of the exchanged field.
For scalar fields the eikonal fails at
eighth order in perturbation theory, when it misses the leading behavior
of the exchange-type diagrams.  In a vector theory the eikonal gets the
exchange
diagrams correctly, but fails by ignoring certain non-exchange graphs which
dominate the asymptotic behavior of the full amplitude.
For spin--2 tensor fields the eikonal captures the leading behavior
of each order in perturbation theory, but the sum of eikonal terms
is subdominant to graphs neglected by the
approximation.  We also comment on the eikonal for Yang-Mills vector exchange,
where the additional complexities of the non-abelian theory may be absorbed
into Regge-type modifications of the gauge boson propagators.

\vfill
\eject

\medskip
\goodbreak
\noindent{\bf I.\quad INTRODUCTION}
\smallskip
\nobreak
The eikonal approximation is a technique for estimating the high energy
behavior of a forward scattering amplitude.  Originally developed for
potential scattering in quantum mechanics$^1$, where one approximates the
classical trajectory corresponding to forward scattering by a
straight line and uses a WKB approximation for the
wavefunction, the eikonal was subsequently
extended to quantum field theory$^2$, where one considers the sum of
all exchange-type Feynman diagrams.  These descriptions are closely related,
as the field theoretic eikonal provides
a way of recovering quantum mechanical potential scattering from a quantum
field theory$^3$.

Recently there has been a revival of
interest in this kinematical regime, as several groups have calculated
the amplitude for forward scattering in quantum gravity at energies of order
the Plank scale.  G.~'t Hooft$^4$ solved the Klein-Gordon equation for one
particle
in the presence of the classical gravitational shock wave (Aichelburg-Sexl
metric$^5$) due to the other particle.  Others have studied the field theoretic
eikonal approximation in the context of string theory$^6$.  These results have
recently been
reproduced by H.~and E.~Verlinde$^7$, who reformulated quantum gravity as a
topological field
theory in this kinematical regime -- a construction which is also available
for quantum electrodynamics$^8$.  Our interest in this subject arose
from the observation that all these calculations are in fact equivalent to the
eikonal approximation$^8$.

Does the eikonal approximation correctly estimate the
true high energy behavior of the full perturbation series?
This may be investigated by comparing the eikonal
result to the actual asymptotic behavior of diagrams at low orders in
perturbation theory, with various spins of the exchanged fields.  Among the
extensive literature on the eikonal approximation, the failure
of the eikonal in scalar field
theory was demonstrated by Tiktopoulos and Treiman as well as by Eichten
and Jackiw$^{9}$.  Our discussion of the eikonal for abelian and Yang-Mills
vector exchange follows the extensive calculations summarized in the book by
H.~Cheng and T.~T.~Wu$^{10}$.  Corrections to the leading tensor eikonal
amplitude have been discussed in a string model
of quantum gravity by Amati, Ciafaloni, and Veneziano$^{11}$.

This paper is organized as follows.  Section II gives a brief summary of the
eikonal approximation.  Sections III -- VI discuss the behavior of the
approximation
in scalar, vector, Yang-Mills, and tensor field theory, and section VII
summarizes the results.

\medskip
\goodbreak
\noindent{\bf II.\quad THE EIKONAL APPROXIMATION}
\smallskip
\nobreak
In diagrammatic terms, the eikonal procedure for calculating a high energy
forward scattering amplitude is as follows$^2$.  Consider the sum of
exchange graphs of Fig.~1, where the exchanged particles may be of arbitrary
spin.  Assume that the dominant routing of the hard external momenta is
along the heavy lines indicated in Fig.~1, while the exchanged particles
carry relatively soft momenta.  Use the eikonal approximation for the
propagators carrying a hard momentum,
$$
  {i \over (p + k)^2 - m^2 + i\epsilon} \approx {i \over 2 p \cdot k + i
  \epsilon}.
$$
Here $p$ is one of the hard external momenta, assumed to be on-shell,
and $k$ is a sum of soft exchanged momenta.  Treat the exchanged propagators
exactly, and in computing the vertex factors, ignore recoil of the hard
momenta.

\vglue 2.0in
\nobreak
\centerline{Figure 1.  Sum of exchange graphs.}

In this approximation the amplitude for each exchange graph is finite.
Summing these eikonal exchange amplitudes to all orders yields
$$\eqalign{
{\cal M}_{\rm eik}^{\rm scalar} &= - 2 i s \int d^2x_\perp e^{-i q_\perp \cdot
x_\perp}
\left[ \exp
\left( -{i g^2 \over 2 s} \int {d^2 k_\perp \over (2 \pi)^2}
{e^{-i k_\perp \cdot x_\perp} \over k^2 + \mu^2}\right)
- 1\right]\cr
{\cal M}_{\rm eik}^{\rm vector} &= - {i s \over 2 m^2} \, \delta_{\lambda_1
\lambda_3} \delta_{\lambda_2 \lambda_4} \, \int d^2 x_\perp
e^{-i q_\perp \cdot x_\perp}
\left[ \exp
\left( - i e^2 \int {d^2 k_\perp \over (2 \pi)^2}
{e^{-i k_\perp \cdot x_\perp} \over k^2 + \mu^2}\right)
- 1\right]\cr
{\cal M}_{\rm eik}^{\rm tensor} &= - 2 i s \int d^2 x_\perp
e^{- i q_\perp \cdot x_\perp}
\left[ \exp
\left( i 4 \pi G s \int {d^2 k_\perp \over (2 \pi)^2}
{e^{-i k_\perp \cdot x_\perp} \over k^2 + \mu^2}\right)
- 1\right].\cr
}\eqno(1)$$
These expressions apply to scalar field theory, fermion-fermion scattering in
spinor electrodynamics, and scalar-scalar scattering by graviton exchange
in a Minkowski background, respectively$^{12}$.  In these
expressions $s$ is the square of the center of mass energy, $q_\perp$ is the
two dimensional transverse momentum transfer, and $\mu$ is the mass of the
exchanged particle, which must be introduced
as an infrared regulator in the electrodynamic and gravitational cases.
The vector eikonal contains fermion helicity conserving $\delta$ functions, and
the fermion mass $m$ through the normalization for the spinors.

\medskip
\goodbreak
\noindent{\bf III.\quad EIKONAL VALIDITY: SCALAR THEORIES}
\smallskip
\nobreak
In scalar theories, the assumption that the eikonal momentum routing gives the
dominant $s\rightarrow \infty$, $t/s \rightarrow 0$ behavior of the exchange
diagrams has been shown to be invalid by Tiktopoulos and Treiman as well as
by Eichten and Jackiw$^{9}$ at eighth order in perturbation theory.
At this order
the double cross graph of Fig.~2 has asymptotic behavior that is a factor
of two larger than what is found with the eikonal approximation.  This is
because there are two distinct routings of hard momenta through the diagram
that make equal contributions to the leading asymptotic behavior, as
illustrated in Fig.~2.  Since both routings send hard momenta through a total
of six propagators, they have the same leading behavior.  We note that an
analytic expression for the true high energy behavior of this diagram is
obtained in Cheng and Wu$^{10}$, appendix C.8, including the factor of two.
The situation is even worse at
higher orders in perturbation theory, where there are
exchange graphs for which non-eikonal routings asymptotically dominate
over the eikonal routing$^{9}$.

\vglue 2.0in
\nobreak
\centerline{Figure 2.  Eikonal and non-eikonal routings in the double cross
diagram.}

\medskip
\goodbreak
\noindent{\bf IV.\quad EIKONAL VALIDITY: ABELIAN VECTOR THEORIES}
\smallskip
\nobreak
The vertex factor for vector exchange gives an effective \hbox{(coupling
constant)$^2 = e^2 s$} for the eikonal routing through the exchange diagrams.
This enhances the eikonal routing over non-eikonal, so that
the eikonal approximation correctly obtains the high energy behavior of the
sum of exchange diagrams when abelian vector particles are exchanged.  This is
a
non-trivial statement, as individual exchange diagrams carry powers of
$\log s$,
while the vector eikonal amplitude in (1) has no $\log s$ dependance.  It turns
out that the logarithms cancel when all $n!$ exchange diagrams at order
$e^{2n}$ are summed, as may be seen occurring in the explicit calculations
of Cheng and Wu$^{10}$.

There is the further question of whether the exchange diagrams are the source
of the true high energy behavior of the perturbation series.  In their
extensive calculations in QED, Cheng and Wu$^{10}$ identify classes of diagrams
that give dominant contributions to the elastic scattering amplitude.
These are diagrams that are related by unitarity to particle production;
they first arise at eighth order as illustrated in Fig.~3.  Cheng and
Wu show that the leading behavior of the sum of these diagrams is
of order $e^8s \log s$, and that this logarithm does not cancel when all
eighth order graphs are summed.  The eikonal approximation has failed, since
the eikonal sum
of four photon exchange diagrams is only of order $e^8 s$, as may
be seen by expanding the exponent in (1).

\vglue 2.0in
\nobreak
\centerline{Figure 3.  Inelastic unitarity diagrams.}

\medskip
\goodbreak
\noindent{\bf V.\quad EIKONAL VALIDITY: YANG-MILLS THEORIES}
\smallskip
\nobreak
Yang-Mills exchange graphs differ from their abelian counterparts by the
presence of
group theoretic factors associated with each diagram.  The failure of the
eikonal to include inelastic unitarity diagrams has already been noted in
the abelian case; here we consider how the non-abelian structure affects the
asymptotic behavior of the exchange graphs.

The eikonal approximation is expected to obtain correctly the asymptotic
behavior
of the Yang-Mills exchange graphs, just as it did in the abelian case.
However the
eikonal amplitude does not take on a simple form similar to the expressions
in (1).  In the abelian
case individual exchange diagrams carried powers of $\log s$, which cancelled
when
all exchange graphs of a given order were summed.  This cancellation does not
occur in Yang-Mills theory due to the group factors associated with the
diagrams.

At fourth order, there are two exchange diagrams with asymptotic amplitudes
given explicitly by the following expressions$^{10}$.
$$\eqalign{
{\cal M}_{\rm \hbox{\sevenrm box}} &= - {g^4 s \left( \log s - i \pi \right)
\over 4 \pi m^2}
\, \psi_3^\dagger T_a T_b \psi_1 \, \psi_4^\dagger T_a T_b \psi_2 \,
\int {d^2 k_\perp \over
(2 \pi)^2} {1 \over k_\perp^2 \left(k_\perp - q_\perp \right)^2}\cr
{\cal M}_{\rm \hbox{\sevenrm crossed box}} &= {g^4 s \log s \over 4 \pi m^2}
\, \psi_3^\dagger T_a T_b \psi_1 \, \psi_4^\dagger T_b T_a \psi_2 \,
\int {d^2 k_\perp \over
(2 \pi)^2} {1 \over k_\perp^2 \left(k_\perp - q_\perp \right)^2}\cr
}$$
Here the $\psi_i$ are the color wavefunctions of the external fermions,
$T_a$ and $T_b$ are group generators, and $q_\perp$ is the
two dimensional transverse momentum transfer.
Note that the $s \log s$ terms do indeed cancel in the abelian theory,
when there is no group factor and these amplitudes are simply
added.

\vglue 2.0in
\nobreak
\centerline{Figure 4.  Box and crossed box exchange graphs in the double line
representation.}

In the non-abelian theory, the $s \log s$ terms do not cancel, as
may be made evident by considering $U(N)$ gauge theory with 't Hooft's
double line representation for the gauge propagators.  In this representation
each gluon line is drawn as an equivalent ``fermion-antifermion'' pair$^{13}$.
The fourth order exchange diagrams are drawn in this representation in Fig.~4.
We see that the crossed box diagram has an internal ``fermion'' loop, and
therefore
carries weight $N$ from summing over the $N$ possible colors.  The straight box
diagram has no such loop, and so the $s \log s$ terms do not cancel.

Extensive calculations in non-abelian gauge theory$^{10}$
show that these surviving powers of $\log s$ may be absorbed by Reggeizing
the gauge bosons, i.e.~by shifting the effective propagator for a gauge boson
$$
{g^2 s \over k^2 + i \epsilon} \rightarrow {g^2 s^{\alpha (k^2)}
\over k^2 + i \epsilon}
$$
where the Regge trajectory is given by
$$
\alpha(k^2) = 1 - {3 g^2 \over 4 \pi} \, \left(k^2 + \mu^2\right)
\int {d^2 q_\perp \over (2 \pi)^2} {1 \over \left(q_\perp^2 + \mu^2\right)
\left(\left(q_\perp - k\right)^2 + \mu^2\right)}
$$
for SU(3) gauge theory.
To fourth order only the box and crossed box exchange graphs have $s \log s$
behavior and contribute to this Reggeization.  At higher orders, more than
just exchange graphs must be included~--~for example, at sixth order,
there are six exchange graphs, but a total of twenty-one diagrams, including
inelastic unitarity diagrams, contribute powers of $\log s$.  All these
logarithms are necessary for Reggeization.  By including the
inelastic unitarity diagrams that the naive eikonal ignores, the amplitude can
be written in the form of an ``extended eikonal formula'' that
gives correct asymptotic behavior and also preserves unitarity to all
orders in perturbation theory.  A full discussion of this formula may
be found in the book by Cheng and Wu$^{10}$.

There is an alternative description of the eikonal approximation that, for
abelian gauge theory, is equivalent to the one outlined in section
II.  In this description$^{3}$, one incoming particle establishes a classical
Coulomb potential in its rest frame.  The quantum mechanical wave equation for
the other particle is then solved in the eikonal approximation, i.e.~the
deflection
of the classical trajectory is ignored and a WKB approximation is used for the
wavefunction.

In Yang-Mills theory this alternative
description of the eikonal breaks down, since it doesn't describe even the
exchange graphs correctly.  Applied to Yang-Mills theory this description would
entail solving classical Wong equations$^{14}$ for one particle in the presence
of a classical static gauge field established by the other particle.   This
doesn't correctly describe the exchange graphs because, as can be seen
in Fig.~4, the color charges of the fermions change each time a gluon is
exchanged.  Neither fermion can be thought of as a source for a static
gauge field, since the internal color degrees of freedom have important
dynamics
even in the eikonal kinematic regime.  Similar  behavior has been noted
for the spin degrees of freedom of a vector meson coupled to an abelian
gauge field$^{15}$, which also fails to eikonalize in the high energy limit.

For gravity, in contrast, we do expect this alternate description of the
eikonal
to apply.  In gravity, it is the initial momentum that is the non-abelian
charge carried by the incoming particles.  In the regime $s \rightarrow
\infty$ with $t/s \rightarrow 0$ the interaction is dominated by
soft graviton exchange.  Then the initial charge stays essentially constant
during the interaction, and it makes sense to say
that the incoming particles establish classical gravitational
potentials.  This is the description of the eikonal that 't Hooft used in his
calculation of the gravitational scattering amplitude at Planckian
energies$^4$, as has been discussed in detail by Kabat and Ortiz$^8$.

\medskip
\goodbreak
\noindent{\bf VI.\quad EIKONAL VALIDITY: GRAVITATION}
\smallskip
\nobreak
With spin--2 exchange, the effective \hbox{(coupling constant)$^2 = G s^2$}
for the eikonal routing through the exchange diagrams.
This is an even stronger enhancement of the eikonal contribution than
occurred in the vector case, and it
seems that order by order in perturbation theory the asymptotic behavior of the
sum of all graphs at order $G^n$ is correctly given by the eikonal's sum
of the $n!$ exchange graphs.

To make this plausible, consider the diagrams of Fig.~5, which are the
gravitational
analogs of the inelastic unitarity diagrams that caused problems for the vector
exchange eikonal in section IV.  The leading behavior of
these diagrams has been evaluated in a string model of quantum gravity
by Amati, Ciafaloni, and Veneziano$^{11}$; they find the asymptotic
behavior of the sum to be $\sim G^3 s^3 \log s$.  This is indeed subdominant
to the eikonal sum of three graviton exchange diagrams, which is of
order $G^3 s^4$ as may be seen by expanding the exponent in (1).

However, summing the eikonal approximation to all orders gives the amplitude
$$
{\cal M}_{\rm eik}^{\rm tensor} = {8 \pi G s^2 \over -t} {\Gamma(1 - i G s)
\over \Gamma(1 + i G s)} \left({4 \mu^2 \over -t}\right)^{- i G s}.
$$
This follows from evaluating the $x_\perp$ and $k_\perp$ integrals in (1) for
small regulating mass $\mu$.
The eikonal amplitude is just the Born amplitude
${8 \pi G s^2 \over -t}$ multiplied by an additional factor ${\Gamma(1 - i G s)
\over \Gamma(1 + i G s)} \left({4 \mu^2 \over -t}\right)^{- i G s}$ which is
a pure phase for real $s$.  A more careful
analysis of the kinematics, taking into account the finite mass of the
incoming particles, would show that
the additional factor actually has poles at small $s$ which correspond to
the Coulomb-like bound states arising from the gravitational interaction of
two particles$^8$.
The important point to note is that summing the leading eikonal amplitudes
gives a result $\sim G s^2 / t$ of the same order as the Born amplitude.

This result is subdominant to graphs neglected by the eikonal approximation,
such as the order $G^3 s^3 \log s$ inelastic unitarity diagrams of Fig.~5.
The sum of exchange
graphs at order $G^n$ has subleading behavior $\sim G^n s^n$ which also
is ignored by the eikonal.  Perhaps the sum of these subleading terms
makes only a small modification to the eikonal amplitude in the
regime of interest ($Gs \approx 1$, ${t \over s} \ll 1$), if for example
their effect is to Reggeize
the exchanged gravitons.  Until this is shown to be the case, the reliability
of the gravitational eikonal is uncertain.

\vglue 2.0in
\nobreak
\centerline{Figure 5.  Graviton inelastic unitarity diagrams.}

\medskip
\goodbreak
\noindent{\bf VII.\quad CONCLUSIONS}
\smallskip
\nobreak
In summary, the eikonal fails to get the right behavior of the scalar
exchange graphs, starting at order $g^8$.  With abelian vector exchange, the
eikonal
gets the sum of exchange graphs correctly, but ignores the asymptotically
dominant inelastic unitarity diagrams.  The Yang-Mills eikonal behaves
much like the abelian case, with the additional effect of the non-abelian
structure
being to Reggeize the gauge propagators.  For tensor exchange, the eikonal
captures the leading behavior of each order in perturbation theory, but the
sum of leading terms is subdominant to terms neglected by the approximation.
The reliability of the eikonal amplitude for gravity is uncertain, unless the
neglected terms can be shown to sum to a harmless form in the
regime of interest.

\medskip
\goodbreak
\centerline{\bf ACKNOWLEDGEMENTS}
\smallskip
\nobreak
The author wishes to thank Roman Jackiw and Miguel Ortiz for their
collaboration
in investigating the eikonal approximation, Hung Cheng for valuable comments on
a preliminary version of the manuscript, and Katie O'Dwyer for drawing the
figures.  This material is based on work
supported by a National Science Foundation graduate fellowship.

\medskip
\goodbreak
\centerline{\bf REFERENCES}
\smallskip
\nobreak
\item {1.}  The quantum mechanical eikonal is described in J.~J.~Sakuri,
{\it Modern Quantum Mechanics} (Addison-Wesley, 1985), section 7.4.
\item {2.}  H. Cheng and T. T. Wu, {\it Phys. Rev. Lett.} {\bf 22}, 666 (1969);
H. Abarbanel and C. Itzykson, {\it Phys. Rev. Lett} {\bf 23}, 53
(1969);  M. Levy and J. Sucher, {\it Phys. Rev.} {\bf 186}, 1656 (1969).
\item {3.}  E. Brezin, C. Itzykson, and J. Zinn-Justin, {\it Phys. Rev.}
{\bf D1}, 2349 (1970); W. Dittrich, {\it Phys. Rev.} {\bf D1}, 3345 (1970).
For a discussion in a gravitational context see Kabat and
Ortiz, Ref.~8.
\item {4.}  G. 't Hooft, {\it Phys. Lett.} {\bf 198B}, 61 (1987).
\item {5.}  P. C. Aichelburg and R. U. Sexl, {\it Gen. Rel. Grav.} {\bf 2},
303 (1971); T. Dray and G. 't Hooft, {\it Nucl. Phys.} {\bf B253}, 173 (1985).
\item {6.}  I. J. Muzinich
and M. Soldate, {\it Phys. Rev.} {\bf D37}, 359 (1988); D. Amati, M. Ciafaloni,
and G. Veneziano, {\it Phys. Lett.} {\bf 197B}, 81 (1987); {\it Int. J. Mod.
Phys.} {\bf A3}, 1615 (1988).  See also H. J. de Vega and N. S\'anchez,
{\it Nucl. Phys.} {\bf B317}, 706 (1989); {\it ibid.} {\bf B317}, 731 (1989);
M. E. V. Costa and H. J. de Vega, {\it Ann. Phys.} {\bf 211}, 223 (1991);
{\it ibid.} {\bf 211}, 235 (1991).
\item {7.}  H. Verlinde and E. Verlinde, {\it Scattering
at Planckian Energies}, Princeton preprint PUPT-1279 (September 1991), to
appear in {\it Nucl. Phys.} {\bf B}.
\item {8.}  R. Jackiw, D. Kabat, and M. Ortiz,
{\it Phys. Lett.} {\bf B277}, 148 (1992); D. Kabat and M. Ortiz, {\it Eikonal
Quantum Gravity and Planckian Scattering}, M.I.T. preprint CTP \#2069 (March
1992), submitted to {Nucl. Phys.} {\bf B}.
\item {9.}  G. Tiktopoulos and S. B. Treiman, {\it Phys. Rev.} {\bf D3}, 1037
(1971); E. Eichten and R. Jackiw, {\it Phys. Rev.} {\bf D4}, 439 (1971).
\item {10.}  H. Cheng and T. T. Wu, {\it Expanding Protons: Scattering at
High Energies} (M.I.T. Press, Cambridge MA 1987), and references therein.
\item {11.}  D. Amati, M. Ciafaloni, and G. Veneziano, {\it Nucl. Phys.}
{\bf B347}, 550 (1990); {\it Planckian Scattering Beyond the Semiclassical
Approximation}, CERN preprint CERN-TH.6395/92 (February 1992).
\item {12.}  The scalar and vector eikonal expressions may be found in
Abarbanel
and Itzykson, Ref.~2, equations (9) and (10).  The gravitational eikonal is
in Kabat and Ortiz, Ref.~8, following equation (2.4).
\item {13.}  The double line representation is described by S. Coleman, {\it
Aspects of Symmetry} (Cambridge University Press, 1985), p. 370.
\item {14.}  M. Carmeli, K. Huleihil, and E. Leibowitz, {\it Gauge Fields:
Classification and Equations of Motion} (World Scientific, 1989).
\item {15.}  H. Cheng and T. T. Wu, {\it Phys. Rev.} {\bf D5}, 445 (1972).

\par
\vfill
\end